\documentclass[a4paper,12pt]{iopart}
\usepackage[]{amssymb}
\usepackage[]{amsthm}
\usepackage[]{mathrsfs}
\usepackage[]{tensor}
\usepackage[symbol]{footmisc}
\usepackage[]{bm}
\usepackage[]{eucal}
\usepackage[unicode]{hyperref}
\usepackage{xcolor}
\definecolor{pastgreen}{HTML}{669900}
\definecolor{pastblue}{HTML}{336699}
\definecolor{linkcol}{HTML}{663333}
\hypersetup{colorlinks,linkcolor={linkcol},citecolor={pastblue},urlcolor={pastgreen}}

\newcommand{\dd}{\partial}
\newcommand{\df}{\mathrm{d}}
\newcommand{\w}{\wedge}
\newcommand{\Lie}{\pounds}
\newcommand{\nab}[1]{\nabla_{\!#1}}
\newcommand{\qqd}{\ , \quad}

\newcommand{\heq}[1]{\buildrel #1 \over =}
\newcommand{\bc}{\begin{center}}
\newcommand{\ec}{\end{center}}
\newcommand{\be}{\begin{equation}}
\newcommand{\ee}{\end{equation}}
\newcommand{\txtand}{\quad \textrm{and} \quad}

\theoremstyle{plain} \newtheorem{tm}{Theorem}[]
\theoremstyle{plain} \newtheorem{lm}[tm]{Lemma}
\theoremstyle{definition} \newtheorem{defn}[tm]{Definition}
\newcommand{\btm}{\begin{tm}}
\newcommand{\etm}{\end{tm}}
\newcommand{\blm}{\begin{lm}}
\newcommand{\elm}{\end{lm}}
\newcommand{\bdefn}{\begin{defn}}
\newcommand{\edefn}{\end{defn}}
\newcommand{\prf}[1]{\noindent \textbf{Proof.} #1 \qed}

\begin{document}

\begin{flushright}
ZTF-EP-15-03
\end{flushright}

\title[Symmetry inheritance of scalar fields]{Symmetry inheritance of scalar fields}

\bigskip

\author{Ivica Smoli\'c}

\address{Department of Physics, Faculty of Science, University of Zagreb, p.p.~331, HR-10002 Zagreb, Croatia}

\eads{\mailto{ismolic@phy.hr}}

\vspace{20pt}

\begin{abstract}
Matter fields don't necessarily have to share the symmetries with the spacetime they live in. When this happens, we speak of the symmetry inheritance of fields. In this paper we classify the obstructions of symmetry inheritance by the scalar fields, both real and complex, and look more closely at the special cases of stationary and axially symmetric spacetimes. Since the symmetry noninheritance is present in the scalar fields of boson stars and may enable the existence of the black hole scalar hair, our results narrow the possible classes of such solutions. Finally, we define and analyse the symmetry noninheritance contributions to Komar mass and angular momentum of the black hole scalar hair.
\end{abstract}


\bigskip


\vspace{20pt}

%%%%%%%%%%%%%%%%%%%%%%%%%%%%%%%%%%%%%%%%%%%%%%%%%%%%%%%%%
%%%%%%%%%%%%%%%%%%%%%%%%%%%%%%%%%%%%%%%%%%%%%%%%%%%%%%%%%
\section{Imperfect accordance of spacetime and fields}%%%
%%%%%%%%%%%%%%%%%%%%%%%%%%%%%%%%%%%%%%%%%%%%%%%%%%%%%%%%%
%%%%%%%%%%%%%%%%%%%%%%%%%%%%%%%%%%%%%%%%%%%%%%%%%%%%%%%%%

The simplest version of a relativistic theory is modeled by matter fields on a fixed, possibly curved spacetime. From the perspective of general theory of relativity, such fields can be considered as mere test fields since they don't participate in gravitational field equations and thus do not affect the spacetime itself. Hence, it doesn't come as a surprise that it is not necessary for such fields to have the same symmetries as the background spacetime. For example, in a typical relativistic course we shall encounter scalar fields (solutions to Klein-Gordon equation) and electromagnetic fields (solutions to Maxwell's equations) which come in various shapes and forms, possessing more or less symmetries, in a sheer contrast with the underlying Minkowski spacetime, which is maximally symmetric.

\bigskip

On the other hand, if we \emph{don't} neglect the backreaction of the matter fields on the spacetime and look at the \emph{exact} solutions to the gravitational field equations, the interplay between the symmetries of the spacetime and the matter fields becomes far less trivial. Let us assume that the spacetime $(M,g_{ab},\psi)$ consists of a (Hausdorff paracompact connected) $D$-dimensional manifold $M$, metric $g_{ab}$ of Lorentzian type and a matter field, generically denoted by $\psi$ (all three smooth). Furthermore, let us assume that this spacetime allows at least one isometry generated by the Killing vector field $\xi^a$,
\be\label{eq:Lieg}
\Lie_\xi g_{ab} = 0 \ .
\ee
If the field $\psi$ shares this symmetry, that is if $\Lie_\xi \psi = 0$ necessarily holds, we shall say that field $\psi$ \emph{inherits} this particular symmetry. Whether the symmetry inheritance happens will depend on the nature of the field $\psi$, as well as the type of the symmetry. 

\bigskip

Earliest analysis of this type was done for the electromagnetic field, concluding that $\Lie_\xi F_{ab} = b\,{*F}_{ab}$, where $b$ is a constant function if $F_{ab}$ is non-null (the electromagnetic field is null if $F_{ab} F^{ab} = F_{ab}{*F}^{ab} = 0$). This was originally proven via Rainich formalism for non-null fields \cite{Woo73a,Woo73b,MzHRS74,MW75}, then generalized to all electromagnetic fields \cite{WY76a,WY76b}, including composite case with ideal fluid. Finally, the proof was significantly simplified via spinor formalism \cite{Tod06} and expanded to spacetimes with black holes. Short overview of the symmetry inheritance for electromagnetic fields and references to known noninheriting examples can be found in \cite{SKMHH}, section 11.1. 

\bigskip

Similar results for the scalar fields are few and far between in literature. After the pioneer work \cite{Hoen78} on real scalar fields with the simplest form of potential, a couple of related discussions \cite{BST10a,BST10b,GJ14b} appeared only recently. These, however, are focused solely on very specific cases of stationary spacetimes and the inheritance of time invariance. One possible strategy for the general analysis of this problem would be to considered it within the context of initial value formulation: It follows from the results by R\'acz \cite{Racz99,Racz01} that symmetry inheritance of the fields at the initial hypersurface implies inheritance throughout the spacetime via the evolution of initial data. Still, there is no guarantee that such initial conditions are necessary, thereby a different approach is needed. The aim of this paper is to fill the gap in the literature and to narrow the window of possible cases of noninheritance. 

\bigskip

Apart from being just a curious formal question, the importance of symmetry inheritance can be seen in the context of black hole no-hair theorems. The absence of black hole scalar hair was originally proved \cite{Chase70,Beken72,BekenStaticHair,BekenStationaryHair,Teitel72a,Teitel72b} for asymptotically flat static and stationary axisymmetric spacetimes with minimally coupled scalar field, but was later considerably generalized \cite{Beken95,Sud95,Zann95,Heus96,Saa96,AB02,Winst05,GJ14a}. All these proofs make a number of subtle but crucial assumptions, most notably, scalar field has to be regular on the horizon and it has to inherit the spacetime symmetries. Over the time various examples of black holes with scalar hair have been found which evade at least some of the assumptions imposed in the no-hair theorems \cite{Beken96}: BBMB solution \cite{BBM,Beken74} is asymptotically flat spherically symmetric black hole with non-minimally coupled scalar field which diverges on the event horizon (alas, this divergence is harmless for test particles that cross it), asymptotically AdS hairy solutions \cite{MTZ04}, hairy black holes in scalar-tensor theories \cite{SZ14}, etc. Example of a black hole with hair whose existence is due to symmetry noninheritance of the complex scalar field appeared only recently \cite{KKRS,HR14a}. As all such solutions will be possible to test in a foreseeable future \cite{Will07,BJLP14}, a better understanding of the geometrical background which allows the presence of the black hole hair and its physical consequences is of utmost importance.

\bigskip

The paper is organized as follows. Section 2 introduces notation, conventions and gathers some auxiliary technical comments. In section 3 we analyse the symmetry inheritance of the real scalar field (for a general class of theories), and in section 4 the complex scalar field (for the simplest type of scalar potential). Section 5 serves to illustrate the results from previous sections on more concrete examples of spacetimes with symmetries. In this context, we look more carefully at the effect of the black hole horizons on the symmetry (non)inheritance and possible black hole scalar hair that might result from it. In the final section we summarize the results and make some comments on open questions.

\bigskip

%%%%%%%%%%%%%%%%%%%%%%%%%%%%%%%%%%%
%%%%%%%%%%%%%%%%%%%%%%%%%%%%%%%%%%%
\section{Technical introduction}%%%
%%%%%%%%%%%%%%%%%%%%%%%%%%%%%%%%%%%
%%%%%%%%%%%%%%%%%%%%%%%%%%%%%%%%%%%

Let us first clear out the details about the notation and some conventions. The metric signature is $(-,+,+,\dots)$ and we use the natural system of units, $c = G = 1$. We shall employ ``abstract index'' notation \cite{Wald} or ``indexless'' notation of differential forms \cite{Heusler} where appropriate. One of the drawbacks of the indexless notation is that the same symbol, such as $X$, may denote a vector field $X^a$ or a 1-form $X_a$, which should be clear from the context. For any symmetric tensor field $S_{ab}$ and a vector field $X^a$ we define the 1-form $S(X)_a \equiv S_{ab} X^b$, so that for example 2-form $X \w S(X)$ corresponds to $X_{[a} S_{b]c} X^c$ in abstract index notation. Generic Killing vector field is denoted by $\xi^a$ and a parameter of its orbit by $\zeta$. A nonempty open set $O_\xi \subset M$, invariant under the action of the Killing vector field $\xi^a$, is referred to as a \emph{$\xi$-open set. In other words, any orbit of $\xi^a$ is either completely contained in $O_\xi$ or disjoint from it.} The Killing horizon generated by the vector field $\xi^a$ is denoted by $H[\xi]$. The domain of outer communications is abbreviated with ``d.o.c.''.

\bigskip

We always assume that the spacetime is a solution to the gravitational field equation of the form 
\be\label{eq:ET}
E_{ab} = 8\pi T_{ab} \ ,
\ee
where $E_{ab}$ is some polynomial in Riemann tensors (e.g.~Lovelock's class of tensors \cite{Love71}), which might contain covariant derivatives and Levi-Civita tensor (such as the case in gravitational Chern-Simons theories \cite{BCPPS11a,BCPPS11b}) and $T_{ab}$ is the energy-momentum tensor of matter fields. For example, in a case of Einstein's gravitational field equation with cosmological constant $\Lambda$ we have
\be\label{eq:Etensor}
E_{ab} = R_{ab} - \frac{1}{2}\,R g_{ab} + \Lambda g_{ab} \ .
\ee
Also, we always assume that the spacetime allows at least one Killing vector field $\xi^a$, with norm $N = \xi^a \xi_a$ (note that $\Lie_\xi N = 0$). Using the facts that $\Lie_\xi R_{abcd} = 0$, $\Lie_\xi \epsilon_{abc\dots} = 0$ and $\Lie_\xi \nab{a} = \nab{a} \Lie_\xi$, from (\ref{eq:Lieg}) and (\ref{eq:ET}) follows \cite{SmoMes} that
\be\label{eq:LieT}
\Lie_\xi T_{ab} = 0 \ .
\ee
This is the ``central formula'' from which we want to extract as much as possible information on symmetry properties of fields. Apart from equation (\ref{eq:LieT}) one can also exploit the properties of the 2-form $\xi \w T(\xi)$. Namely, due to specific form of the energy-momentum tensor for the scalar fields, such a 2-form is convenient for this analysis since it explicitly contains Lie derivatives of the fields. More concretely, we shall refer to the condition
\be\label{eq:GMS}
\xi \w T(\xi) = 0
\ee 
as the \emph{generalized matter staticity} (GMS) condition. The significance of (\ref{eq:GMS}) lies in various generalizations of Lichnerowicz theorem \cite{C73,Heusler}. Namely, in a stationary spacetime (with corresponding Killing vector field $k^a$), solution to Einstein's gravi\-ta\-ti\-o\-nal field equation, the matter staticity condition $k \w T(k) = 0$ is equivalent to Ricci staticity condition $k \w R(k) = 0$. If this spacetime is asymptotically flat with strictly stationary d.o.c.~($k^a k_a < 0$), possibly containing a nondegenerate, nonrotating, not necessarily connected Killing horizon $H[k]$, then the d.o.c.~is static ($k \w\df k = 0$) iff it is Ricci static. The theorem is immediately valid for the vacuum ($T_{ab} = 0$) solutions to Einstein's equation, since all these are Ricci static. 

\bigskip

%%%%%%%%%%%%%%%%%%%%%%%%%%%%%%
%%%%%%%%%%%%%%%%%%%%%%%%%%%%%%
\section{Real scalar field}%%%
%%%%%%%%%%%%%%%%%%%%%%%%%%%%%%
%%%%%%%%%%%%%%%%%%%%%%%%%%%%%%

The energy-momentum tensor for the real scalar field $\phi$ with some general potential $V$ (such that e.g.~mass term is $V_{\mathrm{mass}} = \mu^2\phi^2/2$) is given by
\be\label{eq:TRphi}
T_{ab} = \nab{a}\phi \nab{b}\phi - \left( \frac{1}{2}\,\nab{c}\phi \nabla^c\phi + V(\phi) \right) g_{ab} \ .
\ee
It is convenient to express $V$ with contractions of the energy-momentum tensor,
\be
V(\phi) = -\frac{T}{D} \pm \frac{D-2}{2D} \sqrt{ \frac{D T_{ab}T^{ab} - T^2}{D-1} } \ ,
\ee
where $T \equiv g^{ab} T_{ab}$. The sign ambiguity present in this equation is irrelevant for the conclusion that follows. Using (\ref{eq:LieT}) we have
\be
0 = \Lie_\xi V(\phi) = \frac{\df V(\phi)}{\df\phi}\,\Lie_\xi \phi \ .
\ee
Therefore, at every point where $V'(\phi) \ne 0$ we can conclude that symmetry is inherited, $\Lie_\xi \phi = 0$. At points where $V'(\phi) = 0$ (which includes, for example, the case of a massless real scalar field with identically vanishing potential $V = 0$) we have to use a slightly different approach. In this case it is convenient to write energy-momentum tensor in the following form,  
\be
T_{ab} = \nab{a}\phi \nab{b}\phi + \frac{T + 2V}{D-2}\,g_{ab} \ ,
\ee
and introduce the 1-form
\be
T(\xi)_a \equiv T_{ab}\xi^b = (\Lie_\xi \phi) (\df\phi)_a + \frac{T + 2V}{D-2}\,\xi_a \ .
\ee
From the assumptions about the symmetry and the assumption that $V'(\phi) = 0$ (which implies $\Lie_\xi V = 0$), it follows that
\be\label{eq:xiLieT}
0 = \Lie_\xi T(\xi) = (\Lie_\xi \Lie_\xi \phi)\,\df\phi + (\Lie_\xi \phi)\,\df (\Lie_\xi \phi) \ .
\ee
Another contraction with $\xi^a$ gives us
\be\label{eq:xixiLieT}
0 = i_\xi \Lie_\xi T(\xi) = \Lie_\xi \left( (\Lie_\xi \phi)^2 \right) \ .
\ee
This implies that $\Lie_\xi \phi$ is constant along the orbits of $\xi^a$, say $\Lie_\xi \phi = a$ with $\Lie_\xi a = 0$ (in other words, $\phi$ is a linear function of the Killing parameter $\zeta$). Using this information back in the equation (\ref{eq:xiLieT}) we get that $a\,\df a = 0$ holds at each point of the spacetime. So, assuming that $\df a$ is at least a continuous 1-form within some $\xi$-open set $O_\xi$, it follows that $a$ has to be a constant on $O_\xi$. Let us summarize these conclusions and their immediate consequences in a form of the following theorem.

\btm
Let spacetime $(M,g_{ab},\phi)$ be a solution to the gravitational field equation (\ref{eq:ET}) with the energy-momentum tensor $T_{ab}$ of the form (\ref{eq:TRphi}), allowing a Killing vector field $\xi^a$. Then the symmetry is inherited by the scalar field, $\Lie_\xi \phi = 0$, at each point where $V'(\phi) \ne 0$. On each $\xi$-open set $O_\xi \subset M$ where $V'(\phi) = 0$ and $\Lie_\xi \phi$ is a $C^1$ function, $\df \Lie_\xi \phi = 0$ holds on $O_\xi$. Furthermore, assuming that the orbits of $\xi^a$ don't run into a singularity, the symmetry inheritance $\Lie_\xi \phi = 0$ will hold within $O_\xi$ if these curves are compact (topological circles) or if $\phi$ is bounded on noncompact orbits. Conversely, on any $\xi$-open set $O_\xi \subset M$ where both $\df \Lie_\xi \phi$ and $V'(\phi)\Lie_\xi \phi$ identically vanish, the equation (\ref{eq:LieT}) necessarily holds.
\etm

\bigskip

An example of unbounded massless scalar field, a linear function of time in a static spacetime has been found by Wyman \cite{Wyman81}. In order to make things as simple as possible, Wyman has assumed that both the spacetime and the field $\phi$ are spherically symmetric (we have just seen that the second assumption is in fact superfluous). From here Einstein's gravitational field equations imply that at least one of the derivatives, $\dd_t \phi(t,r)$ or $\dd_r \phi(t,r)$ must vanish. In the latter case, denoted by Wyman as the ``Case II'', there are two solutions of the form
\be\label{eq:WymanII}
\df s^2 = -e^{\nu(r)}\,\df t^2 + e^{\lambda(r)}\,\df r^2 + r^2 \left( \df\theta^2 + \sin^2\theta\,\df\varphi^2 \right) \qqd \phi = \gamma t \ ,
\ee
where $\gamma$ is just a dimensionful conversion factor. A simpler solution is given by $e^\nu = 8\pi\gamma^2 r^2$ and $e^\lambda = 2$, while in the second solution functions $\nu$ and $\lambda$ are only known in a form of Taylor series. What we have proven is that these Wyman's solutions are in fact the ``worst case scenario'' for the symmetry noninheritance of the real scalar field with the energy-momentum tensor (\ref{eq:TRphi}).

\bigskip

There is a generalization of the scalar field Lagrangians proposed by \cite{APDM99} (see also \cite{APMS01}), known as ``k-essence'' theories, which serve as a generic model for the inflationary evolution of the universe. In these theories the scalar field contribution to Lagrangian is proportional to the (sufficiently smooth) function $p = p(\phi,X)$ with $X = -\frac{1}{2}\,\nab{a}\phi \nabla^a\phi$, so that the energy-momentum tensor is given by
\be\label{eq:Tnonc}
T_{ab} = p_{,X} \nab{a}\phi \nab{b}\phi + p g_{ab} \ ,
\ee 
where $p_{,X} \equiv \dd p/\dd X$ (and likewise $p_{,\phi} \equiv \dd p/\dd \phi$). For example, the ``canonical'' scalar field described by (\ref{eq:TRphi}) corresponds to the choice $p = X - V(\phi)$. Other typical members of k-essence theories include ghost condensate model \cite{AHCLM04},
\be
p = -X + \frac{X^2}{M^4} \qqd M = \textrm{const.}
\ee
dilatonic ghost model \cite{PT04},
\be
p = -X + \frac{X^2 e^{\lambda\phi}}{M^4} \qqd M,\lambda = \textrm{const.}
\ee 
and DBI model \cite{ST04,MY08}
\be
p = f(\phi) \left( 1 - \sqrt{1 - \frac{2X}{f(\phi)}} \right) - V(\phi) \ .
\ee
An analysis of symmetry inheritance in k-essence theories has been recently presented in \cite{GJ14b}. This, however, is narrow in scope since it only deals with the noncanonical scalar fields in stationary axisymmetric spacetimes, solutions to Einstein field equations.

\bigskip

\blm
Let spacetime $(M,g_{ab},\phi)$ be a solution to the gravitational field equations (\ref{eq:ET}) with the energy-momentum tensor $T_{ab}$ of the form (\ref{eq:Tnonc}), and allowing a Killing vector field $\xi^a$. Then the functions $p$ and $X$ satisfy
\be\label{eq:Liep}
\Lie_\xi p = p_{,\phi} \Lie_\xi \phi + p_{,X} \Lie_\xi X = 0
\ee
and
\be\label{eq:LieXpX}
\Lie_\xi (Xp_{,X}) = 0
\ee
at each point of the manifold $M$.
\elm

\medskip

\prf{
Using the fact that
\be\label{eq:TXp}
T = g^{ab} T_{ab} = -2X p_{,X} + Dp
\ee
and thus
\be
T_{ab} T^{ab} = D(D-1) p^2 - 2(D-1)Tp + T^2 \ ,
\ee
the function $p$ (with irrelevant sign ambiguity) can be expressed as
\be\label{eq:pTT}
p = \frac{T}{D} \pm \frac{1}{D} \sqrt{\frac{D T_{ab} T^{ab} - T^2}{D-1}} \ .
\ee
Equations (\ref{eq:Liep}) and (\ref{eq:LieXpX}) follow from (\ref{eq:LieT}) and Lie derivatives of (\ref{eq:pTT}) and (\ref{eq:TXp}) with respect to the Killing vector field $\xi^a$.
}

\bigskip

\noindent
We note that this result is already enough to deduce the symmetry inheritance for the canonical scalar field when $V'(\phi) \ne 0$. Namely, in this case we have $p_{,X} = 1$, hence (\ref{eq:LieXpX}) implies that $\Lie_\xi X = 0$ and thus
\be
0 = p_{,\phi} \Lie_\xi \phi + p_{,X} \Lie_\xi X = -V'(\phi) \Lie_\xi \phi \ .
\ee
Before we continue with the analysis, let us introduce two useful terms.

\bdefn
We say that the energy-momentum tensor (\ref{eq:Tnonc}) is \emph{admissible} if the function $p = p(\phi,X)$ is such that $\Lie_v (p_{,X})$ vanishes for any vector field $v^a$ whenever both $X = 0$ and $\Lie_v X = 0$. We say that some point $s \in M$ of the spacetime is \emph{exceptional} (with respect to a Killing vector field $\xi^a$) if all three functions $p_{,\phi}$, $p_{,X}$ and $\Lie_\xi p_{,X}$ vanish at $s$.
\edefn

All the examples of k-essence theories mentioned above have a energy-momentum tensor which is admissible. The exceptional points represent the ``blind spots'' in this approach to symmetry inheritance since at such points all three equations, (\ref{eq:LieT}), (\ref{eq:Liep}) and (\ref{eq:LieXpX}) are automatically satisfied and cannot be used for any further information about $\Lie_\xi \phi$. However, if such points are isolated, conclusions from the neighbourhood can be extended to them using continuity of the fields (note that there are no exceptional points for the canonical scalar field since $p_{,X} = 1$). We shall now present one classification of symmetry inheritance for the noncanonical scalar fields, based upon the fact that $Xp_{,X}$ is constant along the orbits of Killing vector field $\xi^a$ 

\bigskip

\btm\label{tm:nonc}
Let spacetime $(M,g_{ab},\phi)$ be a solution to the gravitational field equations (\ref{eq:ET}) with the (nonsingular) energy-momentum tensor $T_{ab}$ of the form (\ref{eq:Tnonc}), and allowing a Killing vector field $\xi^a$. Then at each point of the spacetime
\begin{itemize}
\item[(a)] $p_{,\phi} \ne 0$ and $Xp_{,X} = 0$ imply $\Lie_\xi \phi = 0$;
\item[(b)] $p_{,\phi} \ne 0$ and $Xp_{,X} \ne 0$ imply $Y \Lie_\xi \phi = 0$, where
\be\label{eq:pphiX}
Y \equiv p_{,\phi} (\Lie_\xi \phi)^2 + 2Xp_{,X}\,\Lie_\xi(\Lie_\xi \phi) \ ;
\ee
\item[(c)] $p_{,\phi} = 0$ and $Xp_{,X} = 0$ with admissible $T_{ab}$ imply $(\Lie_\xi \Lie_\xi \phi)(\Lie_\xi \phi) = 0$, except possibly at points which are exceptional;
\item[(d)] $p_{,\phi} = 0$ and $Xp_{,X} \ne 0$ imply $(\Lie_\xi \Lie_\xi \phi)(\Lie_\xi \phi) = 0$;
\end{itemize}

\medskip

\noindent
Furthermore, let $\gamma$ be a nonsingular orbit of $\xi^a$ which doesn't contain exceptional points and $\Lie_\xi \phi$ a $C^1$ function. Then
\begin{itemize}
\item[(i)] $Xp_{,X} = 0$ on $\gamma$ implies that $\Lie_\xi \Lie_\xi \phi = 0$ holds along $\gamma$ for admissible $T_{ab}$;
\item[(ii)] $Xp_{,X} \ne 0$ on $\gamma$ implies that the function $\Lie_\xi \phi$ is a solution to $Y = 0$ which is either identically zero along $\gamma$, or doesn't have any zeros on $\gamma$. 
\end{itemize}
\etm

\medskip

\prf{
Let us first assume that $p_{,\phi} \ne 0$. If $p_{,X} = 0$ then $\Lie_\xi \phi = 0$ immediately follows from (\ref{eq:Liep}). If $X = 0$ then (\ref{eq:LieXpX}) together with (\ref{eq:Liep}) again implies that $\Lie_\xi \phi = 0$. Suppose now that $Xp_{,X} \ne 0$. Using (\ref{eq:Liep}) the equation (\ref{eq:LieT}) reduces to
\be\label{eq:Mast}
\Lie_\xi (p_{,X}) \nab{a}\phi \nab{b}\phi + 2 p_{,X} \nab{(a}\phi \nab{b)} \Lie_\xi \phi = 0 \ .
\ee
Now, using (\ref{eq:Liep}) and (\ref{eq:LieXpX}) to express $\Lie_\xi (p_{,X})$ and contracting (\ref{eq:Mast}) with $\xi^a \xi^b$ we get $Y \Lie_\xi \phi = 0$.

\medskip

Let us now assume that $p_{,\phi} = 0$. At all non-exceptional points $Xp_{,X} = 0$ implies that $X = 0$. Then (\ref{eq:Liep}) and (\ref{eq:LieXpX}) imply that $p_{,X} \Lie_\xi X = 0$. If $\Lie_\xi X \ne 0$ and $p_{,X} = 0$ then (\ref{eq:Mast}) contracted with $\xi^a\xi^b$ implies that $\Lie_\xi \phi = 0$ (except possibly at exceptional points). If $\Lie_\xi X = 0$, then $\Lie_\xi (p_{,X}) = 0$ for admissible $T_{ab}$, hence (\ref{eq:Mast}) implies that $(\Lie_\xi \Lie_\xi \phi)(\Lie_\xi \phi) = 0$ (except possibly at exceptional points). Finally, $Xp_{,X} \ne 0$ with (\ref{eq:Liep}) and (\ref{eq:LieXpX}) implies $\Lie_\xi X = 0$ and $\Lie_\xi (p_{,X}) = 0$. Using all this in (\ref{eq:Mast}) contracted with $\xi^a \xi^b$ we get $(\Lie_\xi \Lie_\xi \phi)(\Lie_\xi \phi) = 0$.

\medskip

The case (i) in the second part of the theorem is a corollary to the (a) and (c) cases in the first part of the theorem. In addition, since $\Lie_\xi \phi$ is by assumption continuous, it cannot jump from one to the other constant value along $\gamma$. Let us now assume that $Xp_{,X} \ne 0$ holds along $\gamma$ and look more closely at possible subcases. If $p_{,\phi} = 0$ along the $\gamma$, then $\Lie_\xi \Lie_\xi \phi = 0$ holds and $\Lie_\xi \phi$ automatically satisfies $Y = 0$. If $p_{,\phi} \ne 0$ and $\Lie_\xi \phi \ne 0$ along the $\gamma$, then the equation $Y = 0$ holds and can be formally integrated as
\be\label{eq:psipsi}
\frac{1}{\psi(\zeta)} - \frac{1}{\psi(\zeta_0)} = \frac{1}{2Xp_{,X}}\,\int_{\zeta_0}^\zeta p_{,\phi}(z)\,\df z
\ee
with $\psi \equiv \Lie_\xi \phi$. Since by assumption $p_{,\phi} \ne 0$, we have either $p_{,\phi} > 0$ or $p_{,\phi} < 0$ along the $\gamma$ and thus $\psi$ is a monotone function. Can we have $\psi$ which is zero at some points and of the form (\ref{eq:psipsi}) at other points of $\gamma$? Let us denote by $\tilde{\gamma}$ an open connected subset of $\gamma$ where $\psi$ is given by (\ref{eq:psipsi}) and by $n\in\gamma$ a boundary point of $\tilde{\gamma}$ where $\psi = 0$. Since by assumption $\psi$ is at least continuous function, $1/\psi$ would be unbounded in a neighbourhood of $n$, hence $p_{,\phi}$ would be singular (which we discard as unphysical). So, as long as $p_{,\phi}$ doesn't have any zeros on $\gamma$, the function $\psi$ is either zero or given by (\ref{eq:psipsi}) at all points of $\gamma$. What if $p_{,\phi}$ has zeros on $\gamma$, but is not identically zero along $\gamma$? In principle one could have a ``mixture'' of subsets with $\psi = 0$, $\Lie_\xi \psi = 0$ (but $\psi \ne 0$) and $\psi$ which is a nonconstant solution to $Y=0$ given by (\ref{eq:psipsi}). However, we have already seen that the only two types of $\psi$ which could ``coexist'' on a single orbit are the latter two. In conclusion, $\Lie_\xi \phi$ is either identically zero along $\gamma$, or doesn't have any zeros on $\gamma$.
}

\bigskip

The canonical scalar field with $V'(\phi) = 0$, as well as the case when $p = p(X)$ (such as the ghost condensate model) are covered by the (c) and (d) cases of the theorem \ref{tm:nonc}. More generally, second part of the theorem is sufficient for the elimination of the symmetry noninheritance when the orbits of $\xi^a$ are compact or $\phi$ remains bounded on the noncompact orbits. We note in passing that the equation $Y = 0$ is possible to explicitly integrate in the dilatonic ghost model. Namely, here one has $p_{,\phi} = \lambda(Xp_{,X} - p)$, so that $\Lie_\xi (p_{,\phi}) = 0$ and thus $\phi$, a general solution to $Y = 0$, depends logarithmically on Killing parameter $\zeta$ along the (noncompact) orbits of $\xi^a$.

\bigskip

An important question is whether all the components of (\ref{eq:LieT}) have been used in conclusions of the theorem \ref{tm:nonc}, i.e.~in which cases we have a guarantee that (\ref{eq:LieT}) is completely satisfied. This follows, for example, whenever we may deduce that $\Lie_\xi \phi = 0$ on some $\xi$-open set $O_\xi$. Here one immediately has $\Lie_\xi \nab{a}\phi = 0$, $\Lie_\xi X = 0$ and, at least for admissible $T_{ab}$, the equation (\ref{eq:LieXpX}) implies that $\Lie_\xi (p_{,X}) = 0$ on $O_\xi$. Putting all this together allow us to conclude that $\Lie_\xi T_{ab}$ vanishes on $O_\xi$. However, whenever we just know that $\Lie_\xi \phi$ is a nonzero solution to $Y=0$, the consequences of the remaining constraints from (\ref{eq:LieT}) on the general behaviour of the function $\Lie_\xi \phi$ remain obscure prior to specialization to some more concrete symmetries and pertaining coordinate systems.

\bigskip

Finally, let us observe that
\be
\xi \w T(\xi) = p_{,X} (\Lie_\xi \phi)\,\xi \w \df\phi
\ee
implies that the inheriting scalar field necessarily satisfied GMS (\ref{eq:GMS}). For example, inheritance of stationarity implies matter staticity and then, via Einstein's equation and Lichnerowicz theorem, staticity (this was first observed by Heusler in \cite{Heus96}). Vice versa, GMS together with $p_{,X} \ne 0$ and $\xi \w \df\phi \ne 0$ implies symmetry inheritance of the real scalar field. 

\bigskip

Both canonical and noncanonical scalar fields are minimally coupled to gravity. For a non-minimally coupled scalar field things will get ``messier'' and it is not quite clear if it is possible to obtain a similar general conclusion about the symmetry inheritance. We might use the trick \cite{Saa96} which enables us to transform non-minimally to minimally coupled scalar field via conformal transformation $\hat{g}_{ab} = f(\phi) g_{ab}$. However, in order to get $\Lie_\xi \hat{g}_{ab} = 0$ from $\Lie_\xi g_{ab} = 0$ one would need to know that $\Lie_\xi \phi = 0$, but this is something we are trying to prove in the first place. This means that we have to deal with relaxed notion of the symmetry, $\Lie_\xi \hat{g}_{ab} = \beta \hat{g}_{ab}$, generated by the conformal Killing vector field $\xi^a$ and a function $\beta$, which considerably complicates equation (\ref{eq:LieT}). We shall postpone discussion about such generalizations for the future work.

\bigskip

%%%%%%%%%%%%%%%%%%%%%%%%%%%%%%%%%
%%%%%%%%%%%%%%%%%%%%%%%%%%%%%%%%%
\section{Complex scalar field}%%%
%%%%%%%%%%%%%%%%%%%%%%%%%%%%%%%%%
%%%%%%%%%%%%%%%%%%%%%%%%%%%%%%%%%

Let us now turn to complex scalar field $\phi$. There are two conventional parametrizations of such field, ``Cartesian'' $\phi = \rho + i\sigma$ and ``polar'' $\phi = A e^{i\alpha}$, each with its own advantages (e.g.~the drawback of the latter is that $\alpha$ is undefined whenever $A = 0$). For convenience, in this section we shall use the notation
\be
\dot{f} \equiv \Lie_\xi f \qqd \ddot{f} \equiv \Lie_\xi (\Lie_\xi f)
\ee
for all scalar functions $f$. For example,
\be
\Lie_\xi \phi = \dot{\rho} + i\dot{\sigma} = (\dot{A} + iA\dot{\alpha}) e^{i\alpha} \ .
\ee
We shall always assume that the field is not trivial, so that $\rho$, $\sigma$ and $A$ are not \emph{identically} zero. The symmetry inheritance of complex scalar field is equivalent to $\dot{\rho} = \dot{\sigma} = 0$ or $\dot{A} = \dot{\alpha} = 0$. The analysis of the symmetry properties of the complex scalar field will be more feasible if at least one of the Lie derivatives of these components vanish. We shall say that the symmetry is \emph{partially inherited} (at some subset of the spacetime) if at least one of the functions $\{\dot{\rho},\dot{\sigma},\dot{A},\dot{\alpha}\}$ vanish. We shall investigate all such cases in both parametrizations.

\bigskip

The energy-momentum tensor for the complex scalar field is given by
\be\label{eq:TCphi}
T_{ab} = \nab{(a}\phi \nab{b)}\phi^* - \frac{1}{2}\,\Big( \nab{c}\phi\,\nabla^c\phi^* + V \Big) g_{ab}
\ee
with the potential $V = V(\phi^*\phi)$. Unlike the case of real scalar field, here it is no longer possible to simply express $V$ as a function of various contractions of $T_{ab}$, so we need a different approach. To this end, it will be useful to rewrite tensor (\ref{eq:TCphi}) as
\be
T_{ab} = \nab{a}\rho \nab{b}\rho + \nab{a}\sigma \nab{b}\sigma + \frac{T + V}{D-2}\,g_{ab} =
\ee
\be
= \nab{a} A \nab{b} A + A^2 \nab{a}\alpha \nab{b}\alpha + \frac{T + V}{D-2}\,g_{ab} \ ,
\ee
where $T = g^{ab} T_{ab}$. Note that, just as in the case of real scalar field, symmetry inheritance implies GMS (\ref{eq:GMS}) for the complex scalar field. The converse is less trivial and demands a careful choice of additional assumptions.

\bigskip

The basic idea in the analysis that follows is to exploit various contractions of the equation (\ref{eq:LieT}) with Killing vector field $\xi^a$. First of all, we have 
\be\label{eq:LieTxi}
\Lie_\xi T(\xi) = 0 \ ,
\ee
which can be decomposed in two equations, projection along the $\xi^a$,
\be\label{eq:LieTxixi}
0 = i_\xi \Lie_\xi T(\xi) = \Lie_\xi (T_{ab} \xi^a \xi^b)  
\ee
and the wedge product with $\xi_a$,
\be\label{eq:LiexiwTxi}
0 = \xi \w \Lie_\xi T(\xi) = \Lie_\xi (\xi \w T(\xi)) \ .
\ee

\bigskip

%%%%%%%%%%%%%%%%%%%%%%%%%%%%%%%%%%%%%%%%%
%%%%%%%%%%%%%%%%%%%%%%%%%%%%%%%%%%%%%%%%%
\subsection{Cartesian parametrization}%%%
%%%%%%%%%%%%%%%%%%%%%%%%%%%%%%%%%%%%%%%%%
%%%%%%%%%%%%%%%%%%%%%%%%%%%%%%%%%%%%%%%%%

Using the following contraction of energy-momentum tensor,
\be
T_{ab} \xi^a \xi^b = \dot{\rho}^2 + \dot{\sigma}^2 + \frac{T + V}{D-2}\,N
\ee
and (\ref{eq:LieTxixi}) we have
\be
\Lie_\xi \left( \dot{\rho}^2 + \dot{\sigma}^2 + \frac{N}{D-2}\,V \right) = 0 \ .
\ee
Whence, along the orbits of $\xi^a$ we have
\be\label{eq:LieTxixiV}
\dot{\rho}^2 + \dot{\sigma}^2 + \frac{N}{D-2}\,V = \nu
\ee
where $\nu$ is a function such that $\Lie_\xi \nu = 0$. We can gain some information about $\nu$ by imposing additional physical constraints. For example, strong energy condition (see e.g.~\cite{Curiel14,Poisson}) in $D = 4$ demands that inequality
\be\label{eq:SEC}
\left( T_{ab} - \frac{1}{2}\,T g_{ab} \right) v^a v^b \ge 0
\ee
holds for any timelike vector field $v^a$. So, on a domain of spacetime where $\xi^a$ is timelike, using $v^a = \xi^a$ condition (\ref{eq:SEC}) implies $\nu \ge 0$. 

\bigskip

To make further analysis more concrete we shall assume that $D = 4$ and 
\be
V = V_{\mathrm{mass}} \equiv \mu^2 \phi^* \phi = \mu^2 (\rho^2 + \sigma^2) \ ,
\ee 
where $\mu$ is the mass of the scalar field. Then (\ref{eq:LieTxixiV}) reduces to the differential equation
\be\label{eq:cartesianconst}
\dot{\rho}^2 + \dot{\sigma}^2 + \frac{1}{2}\,N \mu^2 (\rho^2 + \sigma^2) = \nu \ .
\ee
If the symmetry is at least partially inherited on some $\xi$-open set $O_\xi \subset M$, such that e.g.~$\dot{\sigma} = 0$ on $O_\xi$, then it is possible to integrate equation (\ref{eq:cartesianconst}) along the orbits of $\xi^a$ in $O_\xi$. Namely, in this case (\ref{eq:cartesianconst}) reduces to
\be\label{eq:oderho}
\dot{\rho}^2 + \kappa \rho^2 = \lambda \qquad \textrm{with} \qquad \kappa = \frac{1}{2}\,N \mu^2 \txtand \lambda = \nu - \kappa \sigma^2 \ ,
\ee
where $\Lie_\xi \kappa = \Lie_\xi \lambda = 0$. Classification of solutions to this nonlinear differential equation was performed in the Appendix. In a nutshell, real nontrivial solutions are denoted as Type I (linear), Type II (oscillatory) and Type III (exponential). The only noninheriting bounded or periodic solution is given by Type II ($\kappa > 0$ and $\lambda > 0$),
\be\label{eq:rhosin}
\rho = \sqrt{\frac{\lambda}{\kappa}} \, \sin\big(\sqrt{\kappa}(\zeta - \zeta_0)\big) \ .
\ee
Now, using equation (\ref{eq:LieTxi}) and (\ref{eq:rhosin}) one arrives at 
\be
(c^2 - s^2)\kappa\,\df\lambda + s^2\lambda\,\df\kappa + 2sc\lambda\sqrt{\kappa} \left( -2\sqrt{\kappa}\,\df\big(\sqrt{\kappa}(\zeta - \zeta_0)\big) + \mu^2\,\xi \right) = 0
\ee
where we have introduced abbreviations
\be
s \equiv \sin\big( \sqrt{\kappa}(\zeta - \zeta_0) \big) \txtand c \equiv \cos\big( \sqrt{\kappa}(\zeta - \zeta_0) \big) \ .
\ee 
In particular, at points where $c = 1$ and $s = 0$ or $c = 0$ and $s = 1$ this implies
\be
\kappa\,\df\lambda = 0 \txtand \lambda\,\df\kappa = 0 \ .
\ee
Since $\lambda \ne 0 \ne \kappa$ we conclude that $\df\lambda = 0$ and $\df\kappa = 0$ are each satisfied at least at one point of the orbit of $\xi^a$. However, as $\lambda$ and $\kappa$ are constant along the orbit and $\Lie_\xi\df = \df\Lie_\xi$, it follows that $\df\lambda = 0 = \df\kappa$ at \emph{all} points of the orbit. Finally, as all this is valid for \emph{any} orbit of $\xi^a$ contained in $O_\xi$, we conclude that $\lambda$ and $\kappa$ must be constant everywhere in $O_\xi$ in order for the solution (\ref{eq:rhosin}) to be consistent with the basic symmetry equation (\ref{eq:LieT}). From here we immediately see that for Type II solution the norm $N = 2\kappa/\mu^2$ must be a positive constant, which is a highly nontrivial restriction: $\xi^a$ has to be a spacelike Killing vector field with a constant norm. Furthermore, using all this back in the equation (\ref{eq:LiexiwTxi}) gives us
\be
\xi \w \df(\zeta - \zeta_0) = 0
\ee
In other words, $\xi^a$ must be also a hypersurface orthogonal vector field! Unfortunately, we were unable to find an example of such a solution or prove that it cannot exist due to strict constraints.

\bigskip

Analogous conclusion can be derived for Type I solutions with $\lambda > 0$ ($\lambda = 0$ corresponds to the inheriting case). For example, for a massless scalar field ($\mu = 0 = \kappa$) the noninheriting solution is given by $\rho = \sqrt{\lambda}\,(\zeta - \zeta_0)$. Equation (\ref{eq:LieTxi}) now reduces to $\dot{\rho}\,\df\dot{\rho} = 0$, which implies that $\lambda$ must be a constant within $O_\xi$. 

\bigskip

In the case when $\dot{\rho} \ne 0$ and $\dot{\sigma} \ne 0$ we can reach for some other assumptions which might simplify the general analysis. For example, if $\sigma = b \rho$ with $\dot{b} = 0$ then (\ref{eq:cartesianconst}) can be written as
\be
\dot{\rho}^2 + \kappa\rho^2 = \frac{\nu}{1 + b^2} \qquad \textrm{with} \qquad \kappa = \frac{1}{2}\,N \mu^2 \ .
\ee
This is again a differential equation of the form (\ref{eq:oderho}) and the classification of its solutions is same as above. If we choose a $b$ which is everywhere constant (so that the phase $\alpha$ is constant), the energy-momentum tensor (\ref{eq:TCphi}) becomes
\be
T_{ab} = (1 + b^2) \nab{a}\rho \nab{b}\rho - \frac{1}{2}\,(1 + b^2) \left( \nab{c}\rho \nabla^c\rho + \mu^2 \rho^2 \right) g_{ab} \ ,
\ee
which is equivalent to the energy-momentum tensor (\ref{eq:TRphi}) of the real scalar field 
\be\label{eq:relphiRC}
\phi_{\mathbb{R}} = \rho \sqrt{1+b^2} \ .
\ee
Using results from the section 3, we immediately have conclusion that $\dot{\rho}$ and $\dot{\sigma}$ are necessarily everywhere constant in this case. Also, we may easily construct such solutions using known ones with real scalar fields. For example, from Wyman's solution (\ref{eq:WymanII}) we have a complex version, one-parameter class of solutions to Einstein-Klein-Gordon equations, (\ref{eq:ET})--(\ref{eq:Etensor}) with $\Lambda = 0$ and $\Box\phi = 0$,
\be\label{eq:CWyman}
\df s^2 = -8\pi \gamma^2 r^2\,\df t^2 + 2\,\df r^2 + r^2 \left( \df\theta^2 + \sin^2\theta\,\df\varphi^2 \right) , \ \phi = \frac{1+ib}{\sqrt{1+b^2}}\,\gamma t \ .
\ee
But what happens if the phase $\alpha$ is \emph{not} constant? Let us consider the symmetry inheritance using parametrization adopted to the amplitude and the phase of the field $\phi$.

\bigskip

%%%%%%%%%%%%%%%%%%%%%%%%%%%%%%%%%%%%%
%%%%%%%%%%%%%%%%%%%%%%%%%%%%%%%%%%%%%
\subsection{Polar parametrization}%%%
%%%%%%%%%%%%%%%%%%%%%%%%%%%%%%%%%%%%%
%%%%%%%%%%%%%%%%%%%%%%%%%%%%%%%%%%%%%

The equation (\ref{eq:LieTxixi}) in the polar parametrization of the complex scalar field implies that
\be
\Lie_\xi \left( \dot{A}^2 + A^2\dot{\alpha}^2 + \frac{N}{D-2}\,V(A^2) \right) = 0 \ .
\ee
Therefore, along the Killing orbits we have
\be\label{eq:polarconst}
\dot{A}^2 + A^2\dot{\alpha}^2 + \frac{N}{D-2}\,V(A^2) = \lambda
\ee
where $\lambda$ is a function such that $\Lie_\xi \lambda = 0$. In $D = 4$ strong energy condition (\ref{eq:SEC}) implies that $\lambda \ge 0$ whenever $\xi^a$ is timelike. Unlike in the case of Cartesian parametrization, amplitude $A$ and phase $\alpha$ do not appear symmetrically in the equation (\ref{eq:polarconst}), so we need to carefully investigate the possible subcases with partial symmetry inheritance.

\bigskip

If we assume that $\dot{A} = 0$ on a $\xi$-open set $O_\xi$ then the equation (\ref{eq:polarconst}) implies that $\ddot{\alpha} = 0$, i.e.~$\alpha$ is a linear function of the Killing parameter $\zeta$. Furthermore, inserting all this into (\ref{eq:LieTxi}) one gets
\be
A^2\dot{\alpha}\,\df\dot{\alpha} = 0 \ .
\ee
So, if the complex field $\phi$ doesn't vanish ($A \ne 0$) on $O_\xi$, then we may conclude that $\dot{\alpha}$ is a constant within $O_\xi$. This type of scalar fields are well known within the context of boson stars. In essence, these are localized solutions to Klein-Gordon-Einstein equations, introduced in the 1960s \cite{BP66,Kaup68,RB69} (for more recent reviews see \cite{SM03,LP12}), which may serve as candidates for compact astrophysical objects, such as black hole mimickers. A more recent example can be found in \cite{HR14a}, mentioned in the introduction. We shall comment all these solutions in greater detail below in the next section.

\bigskip

On the other hand, if we assume that $\dot{\alpha} = 0$ on a $\xi$-open set $O_\xi$, together with $D=4$ and $V = V_{\mathrm{mass}}$, equation (\ref{eq:polarconst}) simplifies to 
\be\label{eq:Aode}
\dot{A}^2 + \kappa A^2 = \lambda \qqd \kappa = \frac{1}{2}\,N \mu^2 \ .
\ee
So, we again arrive at the differential equation of the form (\ref{eq:oderho}), with the same conclusions as earlier. Complexified version of Wyman's solution (\ref{eq:CWyman}) is an example of scalar field with constant phase and Type I amplitude $A$. The only noninheriting case where the amplitude $A$ is bounded or periodic is that of Type II,
\be
A = \sqrt{\frac{\lambda}{\kappa}}\,\sin\big( \sqrt{\kappa}(\zeta - \zeta_0) \big) \ ,
\ee
with constant $\kappa > 0$ and $\lambda > 0$ within some $\xi$-open set $O_\xi$. Just as above, $\xi^a$ would have to be a spacelike hypersurface orthogonal Killing vector with constant norm on $O_\xi$.

\bigskip

The cases which are not covered with the general analysis from this section are those for which no partial inheritance is fulfilled, either in Cartesian or polar parametrization. It is not quite clear how to proceed with the classification of such completely noninheriting fields, so we leave them as a separate class. Let us summarize the conclusions that we have gained about the symmetry inheritance properties of the complex scalar fields.

\btm
Let spacetime $(M,g_{ab},\phi)$ be a solution to the gravitational field equations (\ref{eq:ET}) with the energy-momentum tensor $T_{ab}$ of the form (\ref{eq:TCphi}), allowing a Killing vector field $\xi^a$. Then 
\begin{itemize}
\item[(a)] On any $\xi$-open set $O_\xi \subset M$ where the field $\phi$ is nonvanishing and $\Lie_\xi A = 0$ holds, $\Lie_\xi \alpha$ is necessarily constant; 
\item[(b)] On any $\xi$-open set $O_\xi \subset M$ where $\Lie_\xi \alpha = 0$ holds and $V = V_{\mathrm{mass}}$, the only noninheriting function $A$ which is bounded or periodic along every orbit of the $\xi^a$ in $O_\xi$ is Type II solution to (\ref{eq:Aode}), which exists only if $\xi^a$ is a spacelike hypersurface orthogonal vector field with constant norm in $O_\xi$. The analogous conclusion holds for a pair of functions $(\rho,\sigma)$ if at least one of the functions $(\Lie_\xi \rho, \Lie_\xi \sigma)$ identically vanishes on $O_\xi$. 
\end{itemize}
\etm

\bigskip

Just as in the case of noncanonical scalar fields, one might ask whether all the components of (\ref{eq:LieT}) have been used in the theorem above. Again, except in those cases when we may conclude that the symmetry is inherited, $\Lie_\xi \phi = 0$, there is no guarantee that (\ref{eq:LieT}) is completely satisfied by some noninheriting field. Remaining constraints may be revealed by further analysis tailored to specific symmetries and field equations.

\bigskip

%%%%%%%%%%%%%%%%%%%%%%%%%%%%%%%%%%%%%%%%%%%%%%%%%%%%%%
%%%%%%%%%%%%%%%%%%%%%%%%%%%%%%%%%%%%%%%%%%%%%%%%%%%%%%
\section{Stationary, static and axisymmetric cases}%%%
%%%%%%%%%%%%%%%%%%%%%%%%%%%%%%%%%%%%%%%%%%%%%%%%%%%%%%
%%%%%%%%%%%%%%%%%%%%%%%%%%%%%%%%%%%%%%%%%%%%%%%%%%%%%%

Now we turn to more specific examples, those of frequently analysed spacetime isometries. In the case of stationary spacetime we denote the corresponding Killing vector field with $k^a = (\dd/\dd t)^a$ and in the case axisymmetric spacetime we denote the corresponding Killing vector field with $m^a = (\dd/\dd\varphi)^a$. We shall assume that $m^a$ has compact orbits and that it is spacelike in the d.o.c.~(no closed timelike curves). Note, on the other hand, that $k^a$ can typically change its causal character due to possible presence of ergoregions. In a more specific case of a stationary spacetime with hypersurface orthogonal $k^a$, $k \w \df k = 0$, we say that the spacetime is static. 

\bigskip

If the spacetime is stationary then $\Lie_k A = 0$ implies that $\Lie_k \alpha$ is constant. If the d.o.c.~is strictly stationary (that is, if $k^a k_a < 0$ throughout the d.o.c.), such is the case in static spacetimes, and $V = V_{\mathrm{mass}}$ then $\Lie_k\alpha = 0$ implies that either $\Lie_k A = 0$ or 

\begin{itemize}
\item $A$ is Type I with respect to parameter $t$ for $\mu = 0$, or 

\item $A$ is Type III with respect to parameter $t$ for $\mu > 0$.
\end{itemize}
 
\noindent 
Analogous conclusions follow if we replace $A$ and $\alpha$ in the last sentence with $\rho$ and $\sigma$ or vice versa. Similarly, if the spacetime is axisymmetric (not necessarily stationary) then $\Lie_m A = 0$ implies that $\Lie_m \alpha$ is constant. Furthermore, due to $m^a m_a > 0$, assumption $\Lie_m \alpha = 0$ for $V = V_{\mathrm{mass}}$ implies that either $\Lie_m A = 0$ or $A$ is Type II with respect to parameter $\varphi$ for $\mu > 0$ and hypersurface orthogonal $m^a$ with constant norm (noninheriting Type I solution for $\mu = 0$ is excluded since $m^a$ has compact orbits). Just as above, analogous conclusions follow if we replace $A$ and $\alpha$ in the last sentence with $\rho$ and $\sigma$ or vice versa. Also, for any of these Type II solutions in stationary axisymmetric spacetime we would have $\Lie_k A = 0$ (likewise, $\Lie_k \rho = 0$ or $\Lie_k \sigma = 0$), a conclusion that is not immediate in the inheriting cases when $\Lie_m A = 0$, $\Lie_m \rho = 0$ or $\Lie_m \sigma = 0$.

\bigskip

A concrete example for some of these types of solutions can be found among the boson stars. A typical rotating boson star \cite{SM03,LP12} consists of a stationary axisymmetric metric $g_{ab}$ and a scalar field of the form 
\be
\phi = A(r,\theta) e^{i(\omega t - \mathsf{m}\varphi)} \ ,
\ee
with real constant $\omega = \Lie_k \alpha$ and integer ``rotational quantum number'' $\mathsf{m} = -\Lie_m \alpha$. Evidently, symmetry inheritance is broken in the phase of this scalar field. What we have proven is that, under the given assumption of symmetric amplitude $\Lie_k A = \Lie_m A = 0$, this is the only possible form of the noninheriting phase.

\bigskip

%%%%%%%%%%%%%%%%%%%%%%%%%%%%%%%%%%%%%%%%%%%
%%%%%%%%%%%%%%%%%%%%%%%%%%%%%%%%%%%%%%%%%%%
\subsection{Spacetimes with black holes}%%%
%%%%%%%%%%%%%%%%%%%%%%%%%%%%%%%%%%%%%%%%%%%
%%%%%%%%%%%%%%%%%%%%%%%%%%%%%%%%%%%%%%%%%%%

The presence of a black hole in the spacetime may provide an additional constraint on a scalar field. Namely, on any Killing horizon $H[\xi]$ we have (see \cite{Wald}, chapter 12)
\be
R_{ab} \xi^a \xi^b \heq{H} 0
\ee 
and then, via Einstein's equation (\ref{eq:ET})--(\ref{eq:Etensor}),
\be\label{eq:TxixiH}
T_{ab} \xi^a \xi^b \heq{H} 0 \ .
\ee 
This equality has proven to be useful in proofs of constancy of the electric scalar potential $\Phi$ and the magnetic scalar potential $\Psi$ \cite{C73,Smolic12} on each connected component of the Killing horizon $H[\xi]$. In the case of a massless real scalar field with the canonical energy-momentum tensor, $T_{ab} \xi^a \xi^b = a^2 + (\,\dots)N$ and (\ref{eq:TxixiH}) imply that $a \equiv 0$, so that the symmetry inheritance is necessary in the presence of a Killing horizon $H[\xi]$. For example, if we have a static spacetime with Killing horizon $H[k]$, or even a stationary axisymmetric spacetime with Killing horizon $H[\chi]$ (where $\chi^a$ is a linear combination of stationary Killing vector $k^a$ and axial Killing vectors with compact orbits), then it follows that $\Lie_k \phi = 0$. This means that in the static and stationary axisymmetric cases of Bekenstein's no-hair theorem, assumption about the stationarity (and axial symmetry) of the field $\phi$ is in fact superfluous! The same type of argument can be used if the field $\phi$ is invariant under the action of the Killing vector field $\xi^a$ at least on some subset (e.g.~at ``infinity'') of the domain of spacetime that we are investigating.

\bigskip

For the complex scalar field (\ref{eq:TxixiH}) implies
\be\label{eq:onH}
\Lie_\xi \rho \heq{H} 0 \heq{H} \Lie_\xi \sigma \txtand \Lie_\xi A \heq{H} 0 \heq{H} A \Lie_\xi \alpha \ .
\ee
If in addition $\phi$ is nonvanishing on $H[\xi]$, one may conclude that $\Lie_\xi \alpha = 0$ on $H[\xi]$ as well. For example, if the spacetime is static (so that $k^a k_a < 0$ in d.o.c.)~and contains a static Killing horizon $H[k]$ then, using previous analysis we may deduce

\begin{itemize}
\item if $\phi$ is not zero on $H[k]$ then $\Lie_k A = 0$ implies that $\Lie_k \phi = 0$;

\item if $V = 0$ (massless case) then either $\Lie_k \sigma = 0$ or $\Lie_k \rho = 0$ implies via (\ref{eq:onH}) that $\Lie_k \phi = 0$;
\end{itemize}

\noindent
If a stationary spacetime contains a rotating black hole, its horizon will typically be surrounded with an ergoregion $\mathcal{E}$ where the stationary Killing vector is spacelike, $k^a k_a > 0$. For example, if a 4-dimensional asymptotically flat spacetime is Ricci static but not static, d.o.c.~cannot be strictly stationary (see e.g.~\cite{Heusler}, theorem 8.2), thus there has to be an ergoregion. Within the ergoregion a conclusion from above has to be modified: For $V = V_{\mathrm{mass}} \ne 0$ and $\Lie_k \alpha = 0$ it follows that either $\Lie_k A = 0$ or $A$ is of Type II with respect to the parameter $t$ (analogous conclusion holds if we replace $\alpha$ and $A$ with $\rho$ and $\sigma$ or vice versa). However, in the latter case $k^a$ would have to be hypersurface orthogonal, i.e.~spacetime is static for which, due to Vishveshwara's theorem \cite{Vish68,Heusler} ergosurface coincides with the Killing horizon $H[k]$ and thus there is no ergoregion! In conclusion, for $V = V_{\mathrm{mass}} \ne 0$ the vanishing of either $\Lie_k \alpha$, $\Lie_k \rho$ or $\Lie_k \sigma$ implies that $\Lie_k \phi = 0$, at least within the ergoregion.

\bigskip

As an concrete example, let us again look at the HR solution \cite{HR14a}. This is a stationary axisymmetric spacetime with a Killing horizon $H[\chi]$, where $\chi^a = k^a + \Omega_{\mathrm{H}} m^a$ and constant $\Omega_{\mathrm{H}}$ represents the ``angular velocity'' of the horizon. Their ansatz for the scalar field $\phi$ contains assumption $\Lie_k A = \Lie_m A = 0$, from which we may deduce that $\Lie_k \alpha$ and $\Lie_m \alpha$ are necessarily constant. Furthermore, since $\Lie_\chi A = 0$ and (\ref{eq:onH}) imply $\Lie_\chi \alpha = 0$, our general analysis allows us even to write the relation between these constants,
\be
\Lie_k \alpha + \Omega_{\mathrm{H}} \Lie_m \alpha = 0 \ .
\ee
This reveals the necessity of the choice $w = \Omega_{\mathrm{H}} \mathsf{m}$ in \cite{HR14a} between the ``frequency'' $w = -\Lie_k \alpha$ and the ``azimuthal winding number'' $\mathsf{m} = \Lie_m \alpha$.

\bigskip

%%%%%%%%%%%%%%%%%%%%%%%%%%%%%%%%%%%%%%%%%%%%%%%%%%%%%%%%%%%%%%
%%%%%%%%%%%%%%%%%%%%%%%%%%%%%%%%%%%%%%%%%%%%%%%%%%%%%%%%%%%%%%
\subsection{Komar mass and angular momentum of scalar hair}%%%
%%%%%%%%%%%%%%%%%%%%%%%%%%%%%%%%%%%%%%%%%%%%%%%%%%%%%%%%%%%%%%
%%%%%%%%%%%%%%%%%%%%%%%%%%%%%%%%%%%%%%%%%%%%%%%%%%%%%%%%%%%%%%

Let us look more closely at the situation when black hole has a scalar hair due to symmetry noninheritance. We would like to give a more quantitative description on how this hair may contribute to black hole properties. We shall assume that the spacetime is asymptotically flat solution to Klein-Gordon-Einstein equations (with $\Lambda = 0$). If this spacetime is stationary, then it is possible to introduce Komar mass on a spacelike hypersurface $\Sigma$ (extending from spacelike infinity to horizon section $\mathcal{H} = H \cap \Sigma$) via
\be\label{eq:KomarM}
M - M_{\mathrm{H}} = -\frac{1}{4\pi} \int_\Sigma *R(k) = -2\int_\Sigma *T(k) + \int_\Sigma T\,{*k} \ ,
\ee
where $M$ is the ``global'' Komar mass measured at infinity and $M_{\mathrm{H}}$ is the ``local'' Komar mass measured on the horizon \cite{Komar59,Heusler,Poisson}. Similarly, if the spacetime is axially symmetric (not necessarily stationary) then it is possible to introduce Komar angular momentum via
\be\label{eq:KomarJ}
J - J_{\mathrm{H}} = \frac{1}{8\pi} \int_\Sigma *R(m) = \int_\Sigma *T(m) - \frac{1}{2} \int_\Sigma T\,{*m} \ .
\ee
There is a natural choice of terms in (\ref{eq:KomarM}) and (\ref{eq:KomarJ}) which constitute the symmetry noninheritance (``sni'') contribution to Komar mass and angular momentum,
\be\label{eq:sniM}
\Delta M_{\mathrm{sni}}^{(\phi)} \equiv -2 \left( \int_\Sigma \Lie_k A\,{*\df} A + \int_\Sigma A^2 \Lie_k \alpha\,{*\df}\alpha \right) \ ,
\ee
\be\label{eq:sniJ}
\Delta J_{\mathrm{sni}}^{(\phi)} \equiv \int_\Sigma \Lie_m A\,{*\df} A + \int_\Sigma A^2 \Lie_m \alpha\,{*\df}\alpha \ .
\ee
Obviously, in symmetry inheriting case both of these vanish. Also, as long as the Komar masses $M$, $M_H$ and Komar angular momenta $J$, $J_H$ are well defined (finite), so will be $\Delta M_{\mathrm{sni}}^{(\phi)}$ and $\Delta J_{\mathrm{sni}}^{(\phi)}$. We can gain a better insight in these quantities if we rewrite them in the following way. Assuming that $\Sigma$ is a $t = \textrm{const}.$ hypersurface, then 
\be
{*\df} A\,|_\Sigma = \Lie_k A \, {*\df}t\,|_\Sigma = \frac{\Lie_k A}{k^a k_a}\,{*k}\,|_\Sigma
\ee
and similarly for ${*\df}\alpha$. Using this we may rewrite (\ref{eq:sniM}) and (\ref{eq:sniJ}) as
\be
\Delta M_{\mathrm{sni}}^{(\phi)} = -2 \int_\Sigma \frac{(\Lie_k A)^2 + (A \Lie_k \alpha)^2}{k^a k_a}\,{*k} \ ,
\ee
\be
\Delta J_{\mathrm{sni}}^{(\phi)} = \int_\Sigma \frac{(\Lie_k A)(\Lie_m A) + A^2(\Lie_k \alpha)(\Lie_m \alpha)}{k^a k_a}\,{*k} \ .
\ee
Note that for $k^a k_a < 0$ on $\Sigma$ (strictly stationary d.o.c.)~we have $\Delta M_{\mathrm{sni}}^{(\phi)} \ge 0$. 

\bigskip

The HR solution is an concrete example of such black hole with scalar hair, alas it is only known as a numerical solution and thus unpractical for the evaluation of the integrals given above. However, it is not difficult to see that in this solution's sni scalar hair contributions to Komar mass and angular momentum are related by Smarr-like relation
\be
\Delta M_{\mathrm{sni}}^{(\phi)} = 2 \Omega_{\mathrm{H}} \Delta J_{\mathrm{sni}}^{(\phi)} \ .
\ee
It is important to stress \cite{HR14a,HR14b} that, unlike in the case of the Kerr black hole, mass and angular momentum are not enough to fully specify the HR solution. Additional charge is related to the conserved current 1-form (consequence of a global $U(1)$ symmetry),
\be
j_a = -\frac{i}{2} \left( \phi^* \nab{a} \phi - \phi \nab{a} \phi^* \right) \ ,
\ee
which integrated over $\Sigma$ gives global Noether charge (number of scalar particles)
\be
Q = \int_\Sigma * j = \int_\Sigma A^2\,{*\df}\alpha \ .
\ee
In the HR solution, due to constancy of $\Lie_k \alpha$ and $\Lie_m \alpha$, the sni scalar hair mass and angular momentum are related to the charge $Q$ as follows,
\be
\Delta M_{\mathrm{sni}}^{(\phi)} = -2Q \Lie_k \alpha \qqd \Delta J_{\mathrm{sni}}^{(\phi)} = Q \Lie_m \alpha \ .
\ee
The second relation presented here is identical to the relation between the angular momentum and the scalar charge of the rotating boson stars \cite{SM03,GSG14}, which was used to demonstrate the quantization of angular momentum of these objects (boson stars can be seen as macroscopic quantum states).

\bigskip

%%%%%%%%%%%%%%%%%%%%%%%%%%
%%%%%%%%%%%%%%%%%%%%%%%%%%
\section{Final remarks}%%%
%%%%%%%%%%%%%%%%%%%%%%%%%%
%%%%%%%%%%%%%%%%%%%%%%%%%%

As a result of the discussion in section 3 we see that possible obstructions of symmetry inheritance by a real scalar field $\phi$ are highly narrowed for a wide range of theories (with minimal coupling of $\phi$ to gravity). For the canonical fields the Lie derivative $\Lie_\xi \phi$ is constant which may be nonzero only if potential $V(\phi)$ is constant, the orbits of $\xi^a$ are noncompact and $T_{ab}\xi^a\xi^a$ doesn't vanish on the black hole horizon. Similar constraints were found for more general, noncanonical scalar field in the theorem \ref{tm:nonc}. An example of noninheriting solution is already known in the literature \cite{Wyman81}, but it would be interesting to see if there are others of this type. On the other hand, we have seen that the presence of a Killing horizon in a spacetime may imply the vanishing of $\Lie_\xi \phi$, thereby removing unnecessary assumptions about symmetries of the canonical real scalar fields that where imposed in well-known no-hair theorems.

\bigskip

For a complex scalar field we have proved that partial symmetry inheritance may only appear under very restricted circumstances: $\Lie_\xi A = 0$ implies that $\Lie_\xi\alpha$ is constant, while for $V = V_{\mathrm{mass}}$ either $\Lie_\xi\rho = 0$, $\Lie_\xi\sigma = 0$ or $\Lie_\xi\alpha = 0$, implies that possible symmetry noninheritance falls into one of the types from the Appendix. These are even more restricted if the spacetime contains a Killing horizon, as explained in detail above. In a case when symmetry inheritance is broken, there is a possibility of black hole scalar hair to which we can assign the ``sni'' contribution to Komar mass and angular momentum.  

\bigskip

There is still a lot to investigate about the symmetry inheritance of the scalar fields, so we conclude this discussion with a short list of open questions:

\begin{itemize}
\item[(1)] What are the symmetry inheritance properties of the non-minimally coupled scalar fields?
\item[(2)] Is there an example of solution to Einstein-Klein-Gordon equations with Type II or Type III amplitude of the complex scalar field?
\item[(3)] How to classify ``highly contrived'' cases in which none of the four Lie derivatives, $\Lie_\xi\rho$, $\Lie_\xi\sigma$, $\Lie_\xi A$ and $\Lie_\xi\alpha$, vanish?
\item[(4)] What are the symmetry inheritance properties in spacetimes with both scalar and gauge fields?
\end{itemize}

\bigskip

%%%%%%%%%%%
%%%%%%%%%%%
\ack%%%%%%%
%%%%%%%%%%%
%%%%%%%%%%%

We would like to thank Benjamin Mesi\'c for several discussions during the inception of this work, and especially to Istv\'an R\'acz for invaluable advices and encouragement at critical phases in the preparation of the paper. This work was supported by the grant of University of Zagreb, ``Potpore po podru\v cjima'' (PP1.38).

\vspace{30pt}

%%%%%%%%%%%%%%%%%%
%%%%%%%%%%%%%%%%%%
\appendix%%%%%%%%%
%%%%%%%%%%%%%%%%%%
%%%%%%%%%%%%%%%%%%

%%%%%%%%%%%%%%%%%%%%%%%%%%%%%%%%%%%%%%%%
%%%%%%%%%%%%%%%%%%%%%%%%%%%%%%%%%%%%%%%%
\section{A differential equation}%%%%%%%
%%%%%%%%%%%%%%%%%%%%%%%%%%%%%%%%%%%%%%%%
%%%%%%%%%%%%%%%%%%%%%%%%%%%%%%%%%%%%%%%%

\bigskip

We classify the solutions to the nonlinear ordinary differential equation
\be
(y'(x))^2 + \kappa y(x)^2 = \lambda
\ee
with some real constants $\kappa,\lambda$. First we note that for every solution $y(x)$ the function with opposite sign, $-y(x)$, is also a solution. Here we can distinguish following cases (for convenience, we introduce some special names)

\begin{itemize}
\item[i)] $\kappa = 0$

\begin{itemize}
\item[$\bullet$] $\lambda \ge 0$ (\textbf{Type I});

\medskip

\be
y(x) = \sqrt{\lambda}\,x + C \qqd C = \textrm{const.}
\ee

\medskip

\item[$\bullet$] \ $\lambda < 0$; no real functions
\end{itemize}

\bigskip

\item[ii)] $\kappa > 0$ 

\begin{itemize}
\item[$\bullet$] $\lambda > 0$ (\textbf{Type II});

\medskip

\be
y(x) = \sqrt{\frac{\lambda}{\kappa}} \, \sin\big(\sqrt{\kappa}(x - x_0)\big)
\ee

\medskip

\item[$\bullet$] \ $\lambda = 0$; no real nontrivial functions, so $y(x) = 0$

\item[$\bullet$] \ $\lambda < 0$; no real functions
\end{itemize}

\bigskip

\item[iii)] $\kappa < 0$

\begin{itemize}
\item[$\bullet$] $\lambda > 0$ (\textbf{Type IIIa});

\medskip

\be
y(x) = \sqrt{\frac{\lambda}{|\kappa|}} \, \sinh\big(\sqrt{|\kappa|}(x - x_0)\big)
\ee

\medskip

\item[$\bullet$] $\lambda = 0$ (\textbf{Type IIIb});

\medskip

\be
y(x) = C\,e^{\pm \sqrt{|\kappa|} x} \qqd C = \textrm{const.}
\ee

\medskip

\item[$\bullet$] $\lambda < 0$ (\textbf{Type IIIc});

\medskip

\be
y(x) = \frac{|\lambda|}{2}\,e^{\pm\sqrt{|\kappa|}(x - x_0)} + \frac{1}{2|\kappa|}\,e^{\mp\sqrt{|\kappa|}(x - x_0)}
\ee
\end{itemize}

\end{itemize}

The only nonvanishing bounded (for all $x$) solutions are constant ($\lambda = 0$) Type I and Type II solutions. These solutions are also the only nontrivial periodic ones.

\bigskip

%%%%%%%%%%%%%%%%%%%%%%%%%%%%
%%%%%%%%%%%%%%%%%%%%%%%%%%%%
\section*{References}%%%%%%%
%%%%%%%%%%%%%%%%%%%%%%%%%%%%
%%%%%%%%%%%%%%%%%%%%%%%%%%%%

\bibliographystyle{unsrt}
\bibliography{syinsc}

\end{document}